\documentclass[letterpaper, 10 pt, conference]{ieeeconf}  %
\IEEEoverridecommandlockouts                              %
\usepackage{graphicx}
\usepackage{amsmath}
\usepackage{amsfonts}
\usepackage{amssymb}
\usepackage{mathrsfs}

\title{\LARGE \bf
On global convergence of area-constrained formations \\
of hierarchical multi-agent systems
}

\author{Toshiharu Sugie,  Fei Tong, Brian D. O. Anderson, Zhiyong Sun
\thanks{T. Sugie is with Osaka University, Suita, Osaka, Japan.
        {\tt\small  sugie@jrl.eng.osaka-u.ac.jp}
         }%
\thanks{F. Tong is with Kyoto University, Sakyo-ku, Kyoto, Japan.
    {\tt\small  tongfei0205@gmail.com}
}%
\thanks{B.D.O. Anderson is with School of Automation, Hangzhou
Dianzi University, Hangzhou, 310018, China, and Data61-CSIRO
and Research School of Electrical, Energy and Materials Engineering,
Australian National University, Canberra, ACT 2601, Australia.
  {\tt\small Brian.Anderson@anu.edu.au}
  }%
\thanks{Z. Sun is with Eindhoven University of Technology,  the Netherlands,
 {\tt\small  z.sun@tue.nl}
}
}

\begin{document}

\maketitle
\thispagestyle{empty}
\pagestyle{empty}

\begin{abstract}
This paper is concerned with a formation shaping problem for point agents in a two-dimensional space, where control avoids the possibility of reflection ambiguities. One solution for this type of problems was given first for three or four agents by considering a potential function which consists of both the distance error and the signed area terms.
 Then, by exploiting a hierarchical control strategy with such potential functions, the method was extended to any number of agents recently. However, a specific gain on the signed area term must be employed there, and it does not guarantee the global convergence. To overcome this issue, this paper provides a necessary and sufficient condition for the global convergence, subject to the constraint that the desired formation consists of isosceles triangles only.
This clarifies the admissible range of the gain on the signed area for this case.
In addition, as for formations consisting of arbitrary  triangles, it is shown when high gain on the signed area is admissible for global convergence.
\end{abstract}

\section{INTRODUCTION}

Formation control for multi-agent systems is one of the most actively studied topics due to its potential in various applications and theoretical depth. Surveys of formation control are found in \cite{AYFH2008} and \cite{OPA2015}. According to the sensing capability, most approaches are classified as (a) displacement-based control, and (b) distance-based control (see \cite{OPA2015}). In the case of (a), most of the existing works require that all the different local coordinate systems associated with each agent should be aligned with a global coordinate system. In contrast, in the case of (b), each agent requires only the relative position information in each local coordinate system which may not be aligned with a common orientation. This might be a big advantage in practice from the viewpoint of sensor cost.
Hence, the distance-based control has attracted considerable attention (see references in \cite{KBF2009},
\cite{OPA2015},  \cite{SAS2015}, \cite{SAS2018}).
One major drawback is that there can be many undesirable equilibria. Because of this, it is not trivial to guarantee the convergence to the desired formation from all or almost all initial conditions.

In 2017,  Anderson et al. (\cite{ASSAS2017}) introduced an interesting approach for formation shape control,
which enjoys the above merit of distance-based control. They considered a triangular formation or a formation with four agents, which utilizes a potential function including not only the distance errors but also a signed area term. The resultant controller is able to prevent the occurrence of flip ambiguity only with the relative position measurements in each agents' local coordinate frame. Then, Sugie et al. \cite{SASD2018} proposed a hierarchical control strategy which is applicable to any number of agents based on such potential functions, with one restriction that each triangle in the formation should be equilateral. Recently, Cao et al. \cite{CSAS2019} have extended the results of \cite{SASD2018} to the formation incorporating a group of arbitrary triangles. However, a specific gain on the signed area term must be employed there, and it is not clear whether other gains work or not.
In addition, the  global convergence (as opposed to almost global convergence) is not guaranteed.
In this sense, the problem has not been solved completely yet.

The purpose of this paper is to clarify an essential requirement of the gain
on the signed area in order to achieve the desired formation globally.
To this end, this paper provides a necessary and sufficient condition for
the gain to achieve the global convergence for a specific case, where
each subsystem consists of an isosceles triangle.
Furthermore, it is clarified when high gain is admissible in the case
where the desired formation consists of arbitrary triangles.

\section{PROBLEM SETTING}
Consider the system consists of $n$ agents in 2-dimensional space which are governed by
the equations
\begin{eqnarray}
 \dot{p}_i (t)=  u_i (t) \hspace{5mm}  i \in {\cal V}
\end{eqnarray}
where  $p_i(t) \in {\bf R}^2$ and $u_i(t) \in {\bf R}^2$ are the state and the input of agent $i$, and ${\cal V}:=\{1,2,\cdots,n \}$ denotes the set of all agents. The information exchanged topology between agents is described by an undirected graph
${\cal G}=({\cal V},{\cal E})$, where ${\cal E} \subset{ {\cal V} \times {\cal V}}$ denotes the set of edges. For example, $(i,j)\in {\cal E}$ implies that two agents $i$ and $j$ exchange information with each other.
Agent $i$ detects the relative position of the neighbor agent $j$ in its local coordinate frame, and the control input $u_i(t)$ should be of the form of
\begin{eqnarray*}
u_i(t) := f_{i}((p_i(t)-p_j(t)).
\end{eqnarray*}
Define the collective state of all agents by
\begin{eqnarray*}
p(t) := [p_1^T(t), p_2^T(t), \cdots, p_n^T(t)]^T,
\end{eqnarray*}
and let ${\cal P}$ denote the set of all $p \in {\bf R}^{2n}$ which aims to achieve the desired formation, which is specified up to translation and rotation.  The control objective is to find $u_i(t)~(i=1 \sim n)$ which satisfy
\begin{eqnarray*}
\lim_{t \to \infty} p(t) \in {\cal P}
\end{eqnarray*}

We assume that the graph is a triangulated Laman graph \cite{CBB2017}.
Note that while rotation and translation for the formation are acceptable, the position of two agents cannot be exchanged. In particular,  the flipping (or reflection) of each triangle is not acceptable. More precisely, the set ${\cal P}$ will be described
as set out in detail below:

If three agents $\{i,j,k\}$ satisfy
\[
\{(i,j), (j,k), (k,i) \} \in {\cal E},
\]
they are said to form a clique. The set of all such triples $(i,j,k)$ is denoted by ${\cal C}$.
For each clique ($i,j,k$), we define the signed area $Z_{i,j,k}$ by
\begin{eqnarray}
Z_{i,j,k} & := & \frac{1}{2}\mbox{det}\begin{bmatrix}
     1&1&1 \\
       p_i&p_j&p_k\\
    \end{bmatrix}.
  \end{eqnarray}
 It is easy to see that $|Z_{i,j,k}|$ equals the area of the triangle ($i,j,k$), and $Z_{i,j,k}$ is positive if three agents positions, $p_i$, $p_j$ and $p_k$ are located in a counterclockwise ordering; otherwise it is negative.
Then, ${\cal P}$ consists of $p \in {\bf R}^{2n}$ satisfying the following two conditions.

\begin{itemize}
\item[(A)]
 $\| p_i-p_j \| =d_{ij}^*$, ~~ $ \forall  (i,j) \in {\cal E}$
\item[(B)]
$Z_{i,j,k}=Z_{i,j,k}^{*}$,~~ $ \forall  (i,j,k) \in {\cal C}$
 \end{itemize}

\noindent
where $d_{ij}^{*}$ is the given desired distance between two agents $(i,j)$,
 and $Z_{i,j,k}^{*}$ denotes the given desired signed area.
The condition (B) precludes any flipping ambiguity.

\section{PROPOSED METHOD}
In this section,  a special case for three-agent systems will be analyzed which plays the
essential role for this problem.
Then, the hierarchical control strategy proposed by \cite{SASD2018} is applied to solve
the above problem,
which guarantees the global stability
of the whole system.

\subsection{Analysis for a three-agent case.} 

The following result was known for a two-agent case (see \cite{SASD2018}).

{\bf Lemma 1}~ Suppose the system consists of two agents $i$ and $j$. If $p_i$ is fixed and $p_j$ is governed by
\begin{eqnarray}
&\dot{p_j} =  -\frac{\partial V_{(i,j)} }{\partial{p_j}}  \nonumber \\
&V_{(i,j)}  := \frac{1}{4}(||p_i-p_j||^2-{d_{ij}^{*}}^2)^2,
\nonumber
\end{eqnarray}
then $p_j$ converges and all stable equilibria of $p_j$ satisfy
\begin{eqnarray}
||p_i-p_j||= d_{ij}^{*}
\nonumber
\end{eqnarray}
i.e., $p_j$ converges to a point with the desired distance from $p_i$.

Now consider the  subsystem $<S_{ijk}>$ which consists of agents $i$, $j$ and $k$.
We adopt the following potential function proposed by Anderson et al.  \cite{ASSAS2017}
\begin{eqnarray}
\begin{split}
V_{(i,j,k)} =  \frac{1}{4} \Bigl(&( \| p_i-p_j \|^2-{d_{ij}^*}^2)^2\\
&+(\|p_j-p_k \|^2-{d_{jk}^*}^{2})^2\\
&+\left( \| p_k-p_i \|^2- {d_{ki}^*}^2 \right)^2 \Bigr) \\
&+ \frac{1}{2}K(Z_{i,j,k}- Z_{i,j,k}^{*})^2
\end{split}
\label{eq:V-ijk}
\end{eqnarray}
where $K$ is the control gain to be determined.
Note that it was shown in \cite{ASSAS2017}
that the signed area error term (i.e., $\frac{1}{2}K(Z_{i,j,k}- Z_{i,j,k}^{*})^2$ ) plays the central role in avoiding convergence to a formation which is a flipped version of the desired formation.

We assume that agents $i$ and $j$ are fixed with the desired distance
 (i.e., $\|p_i - p_j \|=d_{ij}^{*}$ ) in $<S_{ijk}>$.
Without loss of generality, we assume $Z_{i,j,k}^{*}\textgreater 0$ and
 \begin{eqnarray}
p_i = \begin{bmatrix}
     -c \\
      0
    \end{bmatrix},
    p_j= \begin{bmatrix}
     c \\
      0
    \end{bmatrix},
   p_k= \begin{bmatrix}
     x \\
      y
    \end{bmatrix},
 p_k^{*} =
\begin{bmatrix}
     a \\
     b
    \end{bmatrix},
 \label{eq:pkxy}
\end{eqnarray}
hold where $c:= \frac{1}{2}d_{ij}^{*} >0$. In the above, $p_k^{*}$ denotes the desired position of agent $k$
with $b>0$

\begin{figure}[tbh]
 \begin{center}
  \includegraphics[width=0.5\linewidth]{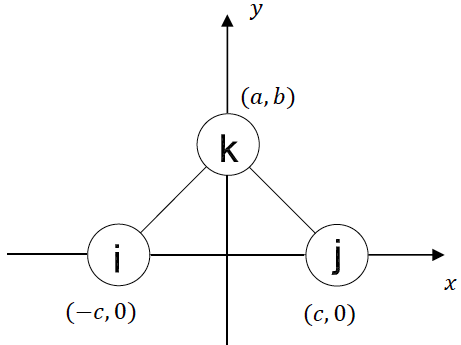}
\end{center}
 \caption{Subsystem $<S_{i,j,k}>$ }
 \label{fig:two}
\end{figure}

Now we obtain the following result in the case of $a=0$ (i.e., $d_{ik}^{*}=d_{jk}^{*}$),
which is a main contribution of this paper (see Fig. \ref{fig:two}).

{\bf Theorem 1:}~
For given $b >0$ and $c>0$, let the position of agents
$i$ and $j$ be  given by (\ref{eq:pkxy}), and assume $d_{ij}^*=2c$ and $d_{ik}^{*}=d_{jk}^{*}=\sqrt{b^2+c^2}$
 in $<S_{ijk}>$. Also assume that agent $k$ is governed by
 \begin{align}
 \dot{p_k} = -\frac{\partial V_{(i, j, k)}}{\partial {p_k}}
\label{eq:pkdot}
\end{align}
while both agents $i$ and $j$ are fixed.
Then,
$p_k$ converges globally to the unique correct equilibrium $p_k^{*} = [0, b]^T$
if and  only if  the following condition holds.
\begin{align}
~~  K >  K_{*},  ~~~   K_{*} := \frac{b^2}{2c^2}
\label{eq:Kstar}
\end{align}

\vspace{2mm}

{\bf Proof}~~
From (\ref{eq:V-ijk}) and (\ref{eq:pkdot}), we have
\begin{align}
\begin{split}
 \dot{p_k} = &-(||p_k-p_j||^2-{d_{kj}^*}^2)(p_k-p_j)\\
& -(||p_k-p_i||^2-{d_{ki}^*}^2)(p_k-p_i)\\
 &-\frac{1}{2}K(Z_{i, j, k}-Z_{i, j, k}^{*})
\begin{bmatrix}
 0&1\\
 -1&0\\
\end{bmatrix}
 (p_i-p_j) \\
& = -
  \begin{bmatrix}
2x(x^2+y^2+2c^2-b^2)\\
2y(x^2+y^2-b^2)+Kc^2(y-b)
\end{bmatrix}
 \end{split}
\end{align}
If the point $[x,y]^T$ is an equilibrium, $\dot{p}_k=0$ must hold. So, we have
\begin{align}
x(x^2+y^2+2c^2-b^2) &=0
\label{eq:equi1}
\\
2y(x^2+y^2-b^2)+Kc^2(y-b) & =0
\label{eq:equi2}
\end{align}

\noindent
{\bf (i)} First we consider the case of $x=0$ in (\ref{eq:equi1}). Then from (\ref{eq:equi2}) we obtain
\begin{align}
(2y^2+2by+Kc^2)(y-b) =0.
\label{eq:equibc}
\end{align}
One real root of the above equation is $y=b$.  Let $\alpha$ and $\beta$ be the roots of
\[
2y^2+2by+Kc^2 = 0,
\]
namely,
\begin{align}
\alpha:= \frac{-b + \sqrt{b^2-2Kc^2}}{2}, ~~
\beta:= \frac{-b - \sqrt{b^2-2Kc^2}}{2}. ~~
\label{eq:beta}
\end{align}
If (\ref{eq:Kstar}) holds, no other real root exists. Hence, no other equilibrium exists.
On the other hand, if
\[
0 < K \leq \frac{b^2}{2 c^2}
\]
holds, $\alpha$ and $\beta$ are two real roots.
All the candidates of the equilibria are given by
\begin{align}
P_a^{*}=
\left[\begin{array}{c}
 0 \\
 b\end{array}\right], ~~
P_b^{*}=
\left[\begin{array}{c}
 0 \\
 \alpha\end{array}\right], ~~
P_c^{*}=
\left[\begin{array}{c}
 0 \\
 \beta\end{array}\right]
\end{align}

From now on, we will check the Hessian at each equilibrium.
Note that Hessian at  point $[x, y]^T$ is given by

\begin{equation}
H=
\begin{bmatrix}
6x^2+2y^2+4c^2-2b^2 & 4xy\\
4xy & 6y^2+2x^2+Kc^2-2b^2
\end{bmatrix}
\nonumber
\end{equation}

\noindent
{\bf (i-a)}~~ The Hessian at point $P_{a}^{*}$ is given by

\begin{align}
H^a =
\begin{bmatrix}
4c^2 & 0\\
0 & 4b^2+Kc^2
\end{bmatrix}
\end{align}
which is positive definite for any $K>0$. Hence $P_a^{*}$ is a stable
equilibrium.

\noindent
{\bf (i-b)}~~ The Hessian at point $P_b^{*}$ is given by
\[
H^b=
\begin{bmatrix}
 2 \alpha^2-2b^2+4c^2   &  0\\
 0  &  b^2-2Kc^2-3b\sqrt{b^2-2Kc^2}
\end{bmatrix}
\]
As for the $(2,2$) entry, it turns out that  $H^b_{22}<0$,  because
the following two relations hold.
\[
\begin{split}
H^b_{22}=b^2 -2Kc^2-3b\sqrt{b^2-2Kc^2}= \\
(\sqrt{b^2-2Kc^2}-3b)\sqrt{b^2-2Kc^2}
\end{split}
\]
and
\[
\sqrt{b^2-2Kc^2}<b<3b
\]
Next consider the $(1,1)$ entry $H^b_{11}$.
This is monotonically increasing with respect to $K$ for $0 < K \leq K_{*}$,
because so is $\alpha^2$. This implies
\[
-2b^2+4c^2 < H^b_{11} \leq -\frac{3}{2}b^2+4c^2.
\]
If
\[
  \frac{b^2}{c^2}  \leq  2
\]
 holds, then $H^b_{11} >0$. So, $P_b^{*}$ is a saddle point.
If
\[
      \frac{8}{3} \leq   \frac{b^2}{c^2}
\]
holds, then $H^b_{11} \leq 0$. Hence, $P_b^{*}$ is an untable equilibrium.
If
\[
2 <\frac{b^2}{c^2} < \frac{8}{3}
\]
the sign of $H_{11}$ depends on $K$.
Since there exists a $K$ at which $H^b_{11}=0$, let $K_0$ be such a $K$, which is calculated as
\begin{align}
K_0= 2  \frac{b}{c}  \sqrt{ \frac{b^2}{c^2}-2 } - 2 \left( \frac{b^2}{c^2} -2 \right)
\label{eq:Kzero}
\end{align}
Hence if $K \leq K_0$ holds, we have $H^b_{11} \leq 0$.  So, $P_b^{*}$ is an unstable equilibrium.
If $K_0 <  K \le K_{*}$ holds,  $H^b_{11}>0$. This implies that $P_b^{*}$ is a saddle point.
%

\noindent
{\bf (i-c)}~~ The Hessian at $P_c^{*}$ is given by
\[
H^c=
\begin{bmatrix}
 2 \beta^2 -2b^2+4c^2   &  0\\
 0  &  b^2-2Kc^2+3b\sqrt{b^2-2Kc^2}
\end{bmatrix}
\]
As long as $0 < K \leq K_{*}$, the $(2,2)$ entry $H^c_{22}$ is nonnegative.
Now we will check the sign of $(1,1)$ entry $H^c_{11}$.
Note that $H^c_{11}$ is monotonically decreasing with respsect to $K$ for
$0 < K \leq K_{*}$, because so is $\beta^2$ from (\ref{eq:beta}). Hence,
\[
-\frac{3}{2}b^2+4c^2 \leq H_{11} < 4c^2
\]
holds. If
\[
 \frac{b^2}{c^2} \leq  \frac{8}{3}
\]
holds, then we have $H^c_{11} \geq 0$. So, $P_{c}^{*}$ is a stable equilibrium.
If
\[
\frac{b^2}{c^2} >   \frac{8}{3},
\]
$H^c_{11}$ can be negative or positive depending on $K$.
Similarly to the argument of $P_{b}^{*}$, it turns out that $H^c_{11}=0$ at $K=K_0$.
So,
if $0 <K \leq K_0$ holds, we have $H^c_{11} \geq 0$.
This implies $P_{c}^{*}$ is a stable equilibrium.
If $K_0 < K \leq  K_{*}$ holds, we obtain $H^c_{11} \leq 0$.
Hence  $P_c^{*}$ is a saddle point.

{\bf (ii)}~  Next, consider the case of
\begin{align}
x^2+y^2+2c^2-b^2=0.
\label{eq:cond2}
\end{align}
Substituting this into (\ref{eq:equi2}), we have
\begin{align}
 c^2\{ (K-4)y-Kb\} =0.
\label{eq:equi22}
\end{align}
When $K=4$ holds, the left hand side of the above equation is positive. So, no equilibrium exists in this case.
Now suppose $ K \neq 4$, then (\ref{eq:equi22}) yields
\begin{eqnarray}
y=\frac{Kb}{K-4}
\label{eq:equiy}
\end{eqnarray}
From (\ref{eq:cond2}), no equilibrium exists when $ \frac{b^2}{c^2} \leq 2$ holds.
Note that the origin, $x=y=0$, cannot be an equilibrium from (\ref{eq:equi22}).
When
\[
 \frac{b^2}{c^2} > 2
\]
holds, $x=\pm \sqrt{b^2-2c^2-y^2}$ is obtained. Hence,
the equilibria are given by
\begin{eqnarray*}
\begin{split}
P_d^{*}&=\begin{bmatrix}
 \sqrt{b^2-2c^2-y^2}\\
 y\\
 \end{bmatrix}\\
P_e^{*}&=\begin{bmatrix}
 -\sqrt{b^2-2c^2-y^2}\\
 y\\
 \end{bmatrix}
 \end{split}
 \end{eqnarray*}
with $y=\frac{Kb}{K-4} $.
Since  $b^2-2c^2-y^2 (=x^2 ) \geq 0$ must hold, $y$ should satisfy
\begin{align}
-\sqrt{b^2-2c^2}\leq y\leq \sqrt{b^2-2c^2}
\label{eq:y-range}
\end{align}
First, we clarify the condition on $K$ to satisfy the above constraints.
If  $K > 4$, we have
\[
y=\frac{Kb}{K-4}=\frac{b}{1-\frac{4}{K}}>b>\sqrt{b^2-2c^2},
\]
which contradicts (\ref{eq:y-range}). Hence
\begin{align}
0< K < 4
\label{eq:Kles4}
\end{align}
should hold. In this case, $y=\frac{Kb}{K-4}$ is negative and (\ref{eq:y-range}) implies
\begin{align}
-\sqrt{b^2-2c^2} \leq \frac{Kb}{K-4},
\end{align}
which reduces to
\begin{align}
K \leq  K_0 
\label{eq:KleqK0}
\end{align}
where $K_0$ is defined by  (\ref{eq:Kzero}).
It can be verified that 
\begin{align}
K_0 \leq K_{*} 
\end{align}
holds  as long as $\frac{b^2}{c^2} \ge 2$ holds.
Hence, if $K > K_{*}$ holds,
either $P_d^{*}$ or $P_e^{*}$ does not exist, because
(\ref{eq:KleqK0}) does not hold in this case.

The Hessian matrix at $P_{d}^{*}$ or $P_{e}^{*} $
is given by
\begin{align}
H^{de}=
\begin{bmatrix}
 4x^2  &  4xy\\
 4xy  &  4y^2+c^2(K-4)\\
\end{bmatrix}
\label{eq:hessian-de}
\end{align}
with $x=\pm \sqrt{b^2-2c^2-y^2}$ and $y=\frac{Kb}{K-4}$.
Simple calculation shows
\[
H^{de}_{11}= 4 x^2>0,~~ \mbox{det} H^{de} =  4x^2c^2(K-4)< 0
\]
hold, because $K < 4$. This implies that
both $P_{d}^{*}$ and $P_{e}^{*} $ are saddle points.

From the arguments of (i) and (ii), the results are summarized as follows:

\begin{itemize}
\item
If $K > K_{*}$, there exists only one equilibrium, $P_a^{*}$, which is stable.
\item
If $K_0 \leq K \leq K_{*}$ and $\frac{b^2}{c^2} \leq  \frac{8}{3}$ hold,
then there exist one saddle point $P_b^{*}$ and
one stable equilibrium $P_c^{*}$, in addition to the stable equilibrium  $P_a^{*}$.
\item
If $K_0 <  K \leq K_{*}$ and $ \frac{8}{3}  < \frac{b^2}{c^2}$ hold,
then there exist one saddle point $P_c^{*}$ and
one unstable equilibrium $P_b^{*}$, in addition to the stable equilibrium  $P_a^{*}$.
\item
If  $0 < K \leq K_0$ and $\frac{b^2}{c^2} \leq 2$ hold,
then there exist one saddle point $P_b^{*}$ and
one unstable equilibrium $P_c^{*}$, in addition to the stable equilibrium  $P_a^{*}$.
\item
If  $0 < K \leq K_0$ and $ 2 < \frac{b^2}{c^2} < \frac{8}{3}$ hold,
then there exist two saddle points  $P_d^{*}$ and $P_e^{*}$,
 and two unstable equilibria $P_b^{*}$ and $P_c^{*}$, in addition to the stable equilibrium  $P_a^{*}$.
\item
If  $0 < K \leq K_0$ and $ \frac{8}{3} \leq \frac{b^2}{c^2}$ hold,
then there exist two saddle points  $P_d^{*}$ and $P_e^{*}$,
one unstable equilibrium $P_b^{*}$, and one stable equilibrium $P_c^{*}$,
in addition to the stable equilibrium  $P_a^{*}$.
\end{itemize}

Hence, $p_k(t)$ converges to the correct point $p_k^{*}$
from any initial condition $p_k(0)$ if and only if $K > K_{*}$ holds.

{\bf (QED)}

As long as the desired formation shape is an isosceles triangle, Theorem 1 gives a necessary and sufficient condition
for global convergence. This is the main contribution of this paper.
The condition $K > K_{*}$ implies that there exists no upper bound on $K$,
which is {\it not} the case for general triangle formations.
We will give a numerical example to show this point later.

In addition, from the proof, we can claim that if
\begin{align}
K \ge 2
\label{eq:Kge2}
\end{align}
holds, $p_k(t)$ converges to the desired point $p_k^{*}$ from {\it almost}
all initial conditions. In fact, when $\frac{8}{3} < \frac{b^2}{c^2}$  holds,
(\ref{eq:Kge2}) means $K > K_0$ because
\[
2-K_0=-2h \sqrt{h^2-2}+2h^2-2 =(h-\sqrt{h^2-2})^2 > 0
\]
holds with $h=\frac{b}{c}$. On the other hand,
when $\frac{8}{3} \geq \frac{b^2}{c^2}$  holds, (\ref{eq:Kge2}) means
$K > K_{*} \geq \frac{4}{3}$.
Note that it is shown that $K=4$ works for achieving {\it almost} global convergence
(for {\it general} triangle formation) in \cite{CSAS2019}.
Our result extends the condition $K=4$ in the case of an isosceles triangle.

\begin{figure}[tbh]
 \begin{center}
  \includegraphics[width=0.9\linewidth]{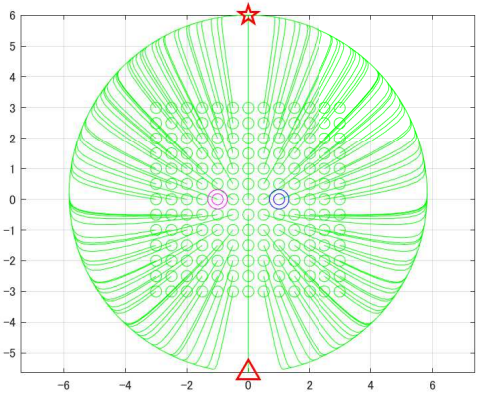}
\end{center}
 \caption{The effect of a saddle point.}
 \label{fig:saddle}
\end{figure}

In order to illustrate the effect of the existence of saddle points, we will show the behavior of
agent $k$ governed by (\ref{eq:pkdot}) with $K=4$ in Fig. \ref{fig:saddle}.
Agents $i$ and $j$ are located at $p_i=[-1, 0]^T$ and $p_j=[1, 0]^T$,
respectively (shown by the circles). The target position of agent $k$ is $p_k^{*}=[0, 6]^T$ (shown by the red star).
Namely, $K_{*}=18$. In this case, one saddle point exists.
Fig. \ref{fig:saddle} shows the trajectories of $p_k(t)$ starting from various points marked at small circles in green.
If the starting point is $[0,y]^T$$(y < 0)$, the trajectory converges
to the wrong point $P_c^{*}=[0, -3-2\sqrt{2}]$ (shown by the red triangle).
The other trajectories converge to the correct point $p_k^{*}$.

\subsection{Extension to general triangle formation}

\begin{figure}[tbh]
 \begin{center}
  \includegraphics[width=0.9\linewidth]{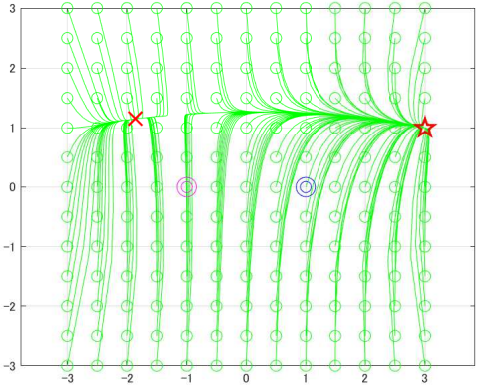}
\end{center}
 \caption{Trajectories of agent $k$ with high gain ($K=80$).}
 \label{fig:incorrect}
\end{figure}

In the case of an isosceles triangle, we have solved the formation problem
as shown in Theorem 1. However, in the case of an arbitrary triangle,
it is known that an incorrect equilibrium may appear
if the gain $K$ is too high. Fig. \ref{fig:incorrect} demonstrates this point,
which shows the trajectory of agent $k$ governed by (\ref{eq:pkdot})
with $K=80$ similar to Fig.  \ref{fig:saddle}.
Agents $i$ and $j$ are pinned at $p_i=[-1,0]^T$ and $p_j=[1,0]^T$
(shown by the circles),
but the target position of agent $k$ is given by  $p_k^{*}=[3,1]^T$
(shown by the red star).
This figure exhibits that there exist two stable equilibria, and some trajectories converge to the wrong point
(shown by the left red cross).
One way to circumvent this problem is to clarify a condition on the triangle shape for which
the global convergence to the correct formation is guaranteed for high gain $K$.
Concerning this issue, the following result is obtained.

{\bf Theorem 2:}~
For given $a$, $b>0$ and $c>0$, let the position of agents
$i$ and $j$ be  pinned at $p_i=[-c,0]^T$ and $p_j=[c,0]^T$,
and the desired position of agent $k$ is given by $p_k^{*}=[a,b]^T$
in $<S_{ijk}>$. Also assume that agent $k$ is governed by
 \begin{align}
 \dot{p_k} = -\frac{\partial V_{(i, j, k)}}{\partial {p_k}}.
\end{align}
When $K \to \infty$,  if
\begin{align}
 \frac{a^2}{c^2} <  8
\label{eq:ac-condition}
\end{align}
holds, $p_k(t)$ converges globally to the unique correct equilibirium $p_k^{*}$.
On the other hand, if
\begin{align}
 \frac{a^2}{c^2} >  8
\label{eq:ac-condition2}
\end{align}
holds, there exists another stable equilibirium. Hence, $p_k(t)$ may converge to an incorrect point depending on the initial condition.

{\bf (Proof)}~~
First, we compute the equilibria candidates. Note that  we have
\begin{align}
\dot{p_k}= -
\begin{bmatrix}
 2x(x^2-a^2-b^2+y^2)+4c^2(x-a)\\
2y(y^2+x^2-a^2-b^2)+Kc^2(y-b)
\end{bmatrix}.
\nonumber
\end{align}
From $\dot{p}_k=0$, we have
\begin{align}
x^2+y^2-a^2-b^2=\frac{2c^2(a-x)}{x}
\nonumber
\end{align}
as long as $x \neq 0$.  Substituting the above into the second entry of
$\dot{p}_k=0$, we obtain
\begin{align}
y=&\frac{bKx}{4a-(4-K)x}.
\nonumber
\end{align}
As $K \to \infty$, $y$ is approaching to $b$.   Hence, substituting $y=b$ into the first entry
of  $\dot{p}_k=0$, we have the following equality.
\begin{align}
(x-a)(x^2+ax+2c^2)=0
\nonumber
\end{align}

{\bf (i)}~ Consider the case of $x=a$. The equilibrium is
$P_a^{*}=[a, b]^T$, and its corresponding Hessian is given by
\begin{align}
H^a=
\begin{bmatrix}
4a^2+4c^2 & 4ab\\
4ab & 4b^2+Kc^2
\end{bmatrix}.
\nonumber
\end{align}
The $(1,1)$ entry $H^a_{11}$ is positive, and the determinant is given by
\begin{align}
\mbox{det} H^a =4Kc^2a^2+4Kc^4+16c^2b^2
\nonumber
\end{align}
which is also positive. Hence $P_a^{*}$ is a stable equilibrium.

{\bf (ii)}~ Next, consider the case of $x^2+ax+2c^2=0$.
If (\ref{eq:ac-condition}) holds, $x^2+ax+2c^2>0$ holds for any $x$.
Hence, there exists no other equilibrium.
Otherwise, let $\alpha$ and $\beta$ be two roots of
$x^2+ax+2c^2=0$, namely
\begin{align}
\alpha=\frac{-a+\sqrt{a^2-8c^2}}{2},~~
\beta=\frac{-a-\sqrt{a^2-8c^2}}{2}.
\nonumber
\end{align}
So, the equiliria are
\begin{align}
P_b^{*} =
\begin{bmatrix}
\frac{-a+\sqrt{a^2-8c^2}}{2} \\
b
\end{bmatrix}, ~~
P_c^{*} =
\begin{bmatrix}
 \frac{-a-\sqrt{a^2-8c^2}}{2}\\
b\\
\end{bmatrix}
\nonumber
\end{align}
The Hessian at $P_b^{*}$ is given by $H^b$ whose ($i, j$) entries are
shown as follows:
\begin{align}
H^b_{11} &= a^2-3a\sqrt{a^2-8c^2}-8c^2 \nonumber \\
H^b_{12} &=H^b_{21}= -2ab+2b\sqrt{a^2-8c^2} \nonumber \\
H^b_{22} &= Kc^2-a^2-a\sqrt{a^2-8c^2}+4b^2-4c^2 \nonumber
\end{align}
It is verified that $H^b_{11} > 0$ holds if $ \frac{a}{c} < -\sqrt{8}$. Because,  $H^b_{11}$ is transformed  to
\begin{align}
H^b_{11}= \sqrt{z^2-8}(\sqrt{z^2-8}-3z) ~~ z:=\frac{a}{c},
\nonumber
\end{align}
and $\sqrt{z^2-8}-3z >0 $ holds from the fact that $(3z)^2 > (z^2-8)$ always holds.
Also, in this case,  from
\[
\det H^b = H^b_{11}H^b_{22} - H^b_{12}H^b_{21},
\]
we have $\det H^b > 0$  when $K \to \infty$.
This implies $H^b$ is positive definite.
Hence, $P_b^{*}$ is a stable (but incorrect)  equilibrium.
On the other hand, the Hessian at $P_c^{*}$ is given by $H^c$ whose ($i, j$) entries are given
as follows:
\begin{align}
H^c_{11} &= a^2+3a\sqrt{a^2-8c^2}-8c^2 \nonumber \\
H^c_{12} &=H^c_{21}= -2ab+2b\sqrt{a^2-8c^2} \nonumber \\
H^c_{22} &= Kc^2-a^2+a \sqrt{a^2-8c^2}+4b^2-4c^2 \nonumber
\end{align}
By a similar argument, it turns out that $H^c$ becomes positive definite if
$\frac{a}{c} > \sqrt{8}$.
So, $P_c^{*}$ is an incorrect stable equiribrium.

In short, if $\frac{a^2}{c^2} > 8$ holds, there always exists a stable equilibrium ($P_b^{*}$ or $P_c^{*}$)
which corresponds to an incorrect point.

{\bf (QED)}

From this theorem, we can use high gain $K$ safely as long as (\ref{eq:ac-condition})
holds even in the  general triangular case.

\subsection{Hierarchical control strategy}

\begin{figure}[htb]
 \begin{center}
  \includegraphics[width=0.5\linewidth]{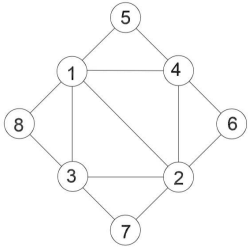}
\end{center}
 \caption{An example of graph structure of a multi-agent system}
 \label{fig:one}
\end{figure}

In this section, we consider the multi-agent system $<S>$ whose graph structure is shown in Fig. \ref{fig:one}
as an example.
The structure is triangulated with all isosceles triangles.
For this system, we apply the hierarchical control strategy \cite{SASD2018}
which will be described below.
Define the control input as
\begin{eqnarray}
 u_i = -\frac{\partial V_{i}}{\partial {p_i}}
\label{eq:u-i}
\end{eqnarray}
where $V_i$ is chosen as shown below.

We show how to choose $V_i$  for the system shown in Fig. \ref{fig:one}.
Consistently with Fig. \ref{fig:one}, we choose $V_i$ for each agent as follows,
where we specify $Z_{i,j,k}^{*} > 0$ for any $(i,j,k) \in {\cal C}$.

Layer 1: ~agent 1,   (which is stationary)
\[
V_1 \equiv 0
\]

Layer 2: ~agent 2 (which is to be a fixed distance from agent 1)
\[
V_2 = V_{(1,2)}
\]

Layer 3: ~agents \{ 3,  4 \}
\[
V_3= V_{(2,1,3)}, ~~ V_4=V_{(1,2,4)}
\]

Layer 4: ~ agents $\{5,6,7,8 \}$
\[
V_5=V_{(1,4,5)}, ~V_6=V_{(4,2,6)},~V_7=V_{(2,3,7)},~V_8=V_{(3,1,8)}
\]

Note that the upper layer agents are never affected by any lower layer agents.
According to the arguments of Theorem 2 of \cite{CSAS2019}, the global asymptotic stability of the whole system
is guaranteed with any gain $K$ satifying (\ref{eq:Kstar}).

{\it Remark:}~Though this paper does not consider collisions between the agents explicitly,
it would be easy to avoid the collisions by applying appropriate existing methods such as
Control Barrier Functions \cite{CBF2019} or Decentralized Navigation Functions \cite{Dimarogonas2016}.

\section{Simulation}

\begin{figure}[bth]
 \begin{center}
  \includegraphics[width=0.8\linewidth]{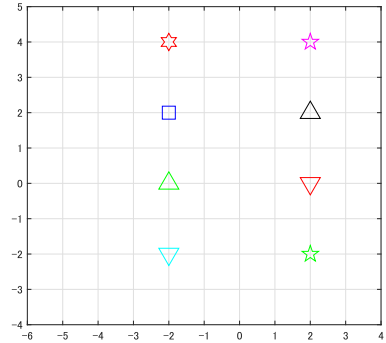}
\end{center}
\caption{Location of 8 agents~($t=0$)}
\label{fig:sim0}
\end{figure}

\begin{figure}[bht]
 \begin{center}
  \includegraphics[width=0.8\linewidth]{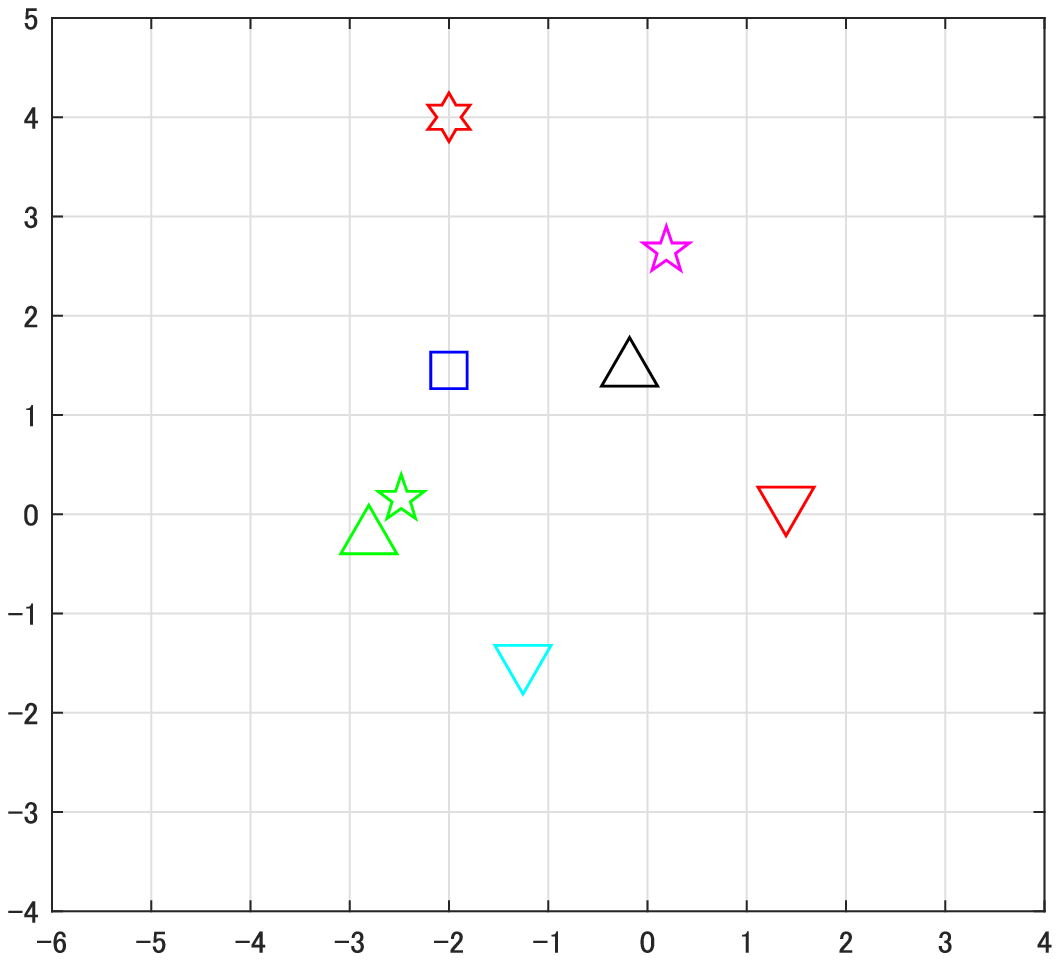}
\end{center}
\caption{Location of 8 agents~($t=0.08$)}
\label{fig:sim1}
\end{figure}

\begin{figure}[htb]
 \begin{center}
  \includegraphics[width=0.8\linewidth]{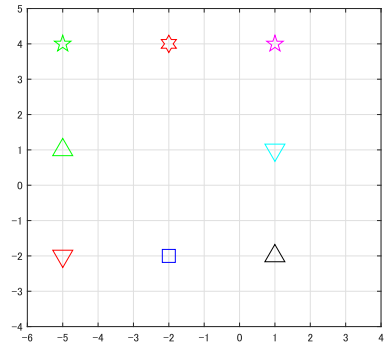}
\end{center}
\caption{Location of 8 agents~($t=0.4$)}
\label{fig:sim3}
\end{figure}

This section shows some simulation results to demonstrate the effectiveness of the proposed method.
The multi-agent system is exactly the same as that in the previous section.
The desired formation also looks like Fig. \ref{fig:one}. Namely,
\begin{align}
d_{12}^{*}=6,~~d_{13}^{*}=d_{14}^{*}=d_{23}^{*}=d_{24}^{*}=3\sqrt{2}
\nonumber
\end{align}
\begin{align}
d_{15}^{*}=d_{18}^{*}=d_{26}^{*}=d_{27}^{*}=d_{37}^{*}=d_{38}^{*}=d_{45}^{*}=d_{46}^{*}=3
\nonumber
\end{align}
\begin{align}
Z_{213}^{*}=Z_{124}^{*}=9, ~~
Z_{145}^{*}=Z_{426}^{*}=Z_{237}^{*}=Z_{318}^{*}=\frac{9}{2}
\nonumber
\end{align}
Eight agents are located initially as shown in Fig. \ref{fig:sim0}. In the left column, agents $1 \sim 4$
are located from top to bottom. Agents $5 \sim 8$  are located similarly in the right column.
Figs. \ref{fig:sim1} $\sim$ \ref{fig:sim3} show snapshots of their locations at $t=0.04 \sim 0.4$, respectively.
It is verified that the correct formation is achieved.

\begin{figure}[tbh]
 \begin{center}
  \includegraphics[width=0.8\linewidth]{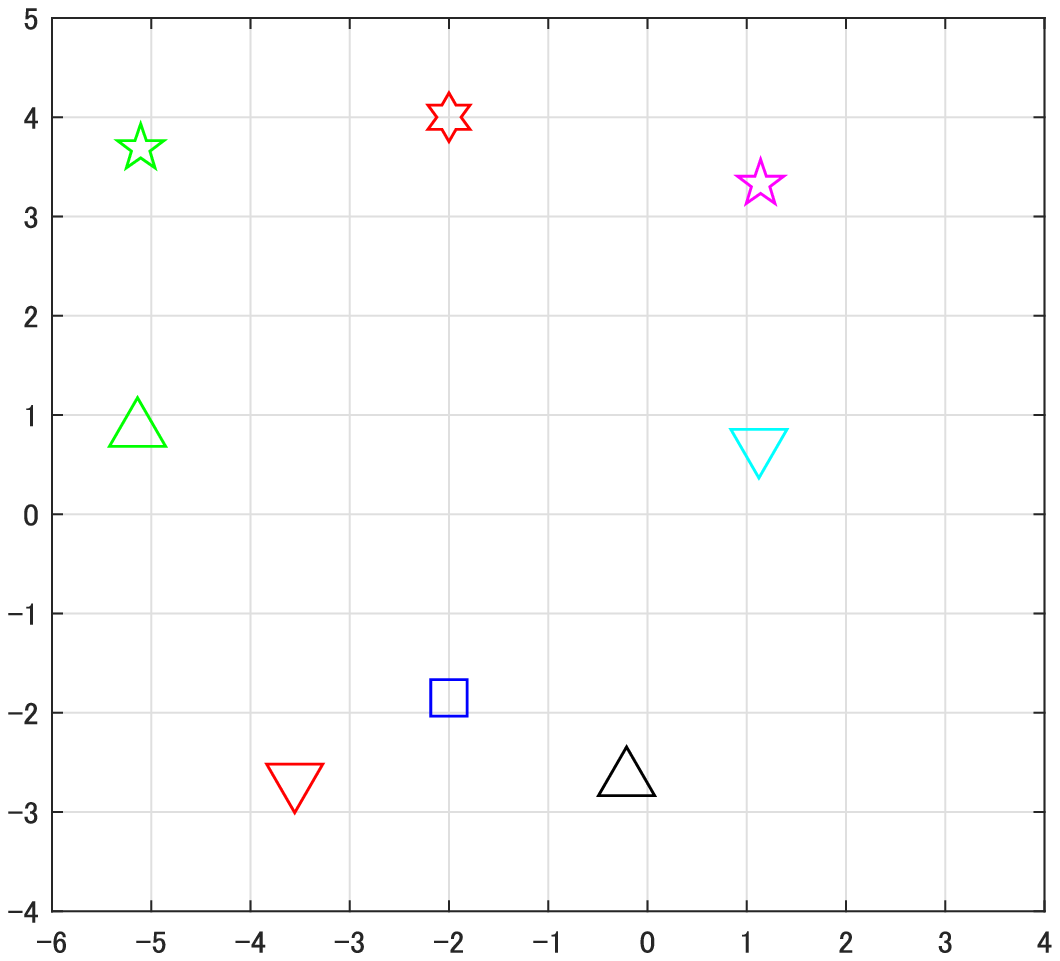}
\end{center}
\caption{Location of 8 agents~~($t=0.16$)}
\label{fig:sim2}
\end{figure}

\section{CONCLUSIONS}

In this paper, we have derived a global convergence condition for the formation shape control of a triangulated Laman graph without flipping in the case where all triangles are isosceles ones. In addition, an almost global convergence condition is clarified. Furthermore, for a formation consisting of general triangles, it is clarified when high gain is admissible for global convergence.

\begin{center}
{\bf Acknowledgements}
\end{center}

This work is supported by JSPS KAKENHI Grant number JP17H03281.
The work of B.D.O. Anderson is supported  by Data-61 CISRO, and by
 the Australian Research Council's Discovery Projects DP-160104500 and DP-190100887.

\end{document}